\documentclass[prl,twocolumn,showpacs]{revtex4-1}
\usepackage{graphicx}
\usepackage{times}
\usepackage{bm}
\usepackage{amsmath, verbatim}

\def\lsim{\mathrel{\rlap{\lower4pt\hbox{\hskip1pt$\sim$}}
    \raise1pt\hbox{$<$}}}                % less than or approx. symbol
\def\gsim{\mathrel{\rlap{\lower4pt\hbox{\hskip1pt$\sim$}}
    \raise1pt\hbox{$>$}}}                % greater than or approx. symbol

\def\be{\begin{equation}}
\def\ee{\end{equation}}
\def\bea{\begin{eqnarray}}
\def\eea{\end{eqnarray}}
\def\bse{\begin{subequations}}
\def\ese{\end{subequations}}

\def\be{\begin{eqnarray}}
\def\ee{\end{eqnarray}}

\newcommand{\expect}[1]{\langle #1 \rangle}

\begin{document}

\title{
Diamagnetic susceptibility obtained from the six-vertex model and its
    implications for the high-temperature diamagnetic state of cuprate
    superconductors}
\author{Jay D. Sau$^{1}$}
\thanks{Present Address: Department of Physics, Harvard University, Cambridge, MA 02138.}
\author{Sumanta Tewari$^{2}$}
%\author{S. Das Sarma$^1$}
\affiliation{$^1$Condensed Matter Theory Center and Joint Quantum Institute, Department of Physics, University of
Maryland, College Park, Maryland 20742-4111, USA\\
$^2$Department of Physics and Astronomy, Clemson University, Clemson, SC
29634}

\begin{abstract}
We study the diamagnetism of the
$6$-vertex model with the arrows
 as directed bond currents.
 To our knowledge, this is the first study of the diamagnetism of this
model.
 %on a 2D square lattice.
 A special version of this model, called F model,
 describes
  the thermal disordering transition of an orbital antiferromagnet, known as $d$-density wave (DDW), a proposed
   state for the pseudogap phase of the high-$T_c$ cuprates.
  We find that the F model is strongly diamagnetic and the susceptibility may diverge in the high temperature
 critical phase with power law arrow correlations.
% One motivation for addressing this problem
These results may explain the surprising recent observation of a
 diverging low-field diamagnetic susceptibility
seen in some optimally doped cuprates within the DDW model of the
 pseudogap phase.

\end{abstract}

\maketitle

%%%%%%%%%%%%%%%%%%%%%%%%%%%%%%%%%%%%%%%%%%%%%%%%%%%%%%%%%%%%%%%%%%%%%

\paragraph{Introduction.}

%%%%%%%%%%%%%%%%%%%%%%%%%%%%%%%%%%%%%%%%%%%%%%%%%%%%%%%%%%%%%%%%%%%%

Experiments on the normal state properties of the cuprate superconductors
continue to pose new theoretical challenges. Above the superconducting transition temperature $T_c$, the cuprates in the underdoped regime
evince a $d$-wave-like gap even in the absence of superconductivity.
 The nature of the system in this pseudogap phase is believed to hold the key \cite{Norman}
to the physics of the high transition temperature itself.
A recent remarkable set of experiments \cite{ong,kivelson} have found evidence of enhanced diamagnetism in the pseudogap phase above $T_c$ at a
doping range near and below the optimal doping.
 In particular, these experiments have revealed that, near optimal doping, the low-field diamagnetic susceptibility $\chi$ \emph{diverges} above $T_c$ as an inverse power of
the applied field $H$, $\chi\sim -H^{-x}$. Here $x$ is a $T$-dependent exponent. The divergence of $\chi$ above $T_c$
 implies underlying critical correlations in an entire phase above $T_c$ \cite{ong} which is not easy to explain by any existing
 theories of the pseudogap phase \cite{ong,kivelson}.
 In this paper, we will address this question within the framework of the $d$-density wave (DDW) state \cite{Nayak}, which was proposed \cite{Chakravarty01} as a candidate state responsible for the
  many anomalous properties of the pseudogap phase.
Our results on diamagnetism will also be important in light of the
recent experiments in Ref.~\cite{armitage} which point towards an
alternative source different from vortices to explain the large
diamagnetism observed in the pseudogap state of the cuprates.

The ordered DDW state consists of counter propagating bond currents on the neighboring plaquettes of a 2D square lattice (Fig.~1) \cite{Nayak}, which can be taken as the Cu lattice of the high-$T_c$ cuprates \cite{Chakravarty01}. The diamagnetism of this state has already been examined within a mean field description \cite{ghosal} in which the direction of the currents on the bonds remains frozen.
%in which the direction fluctuations of the bond currents are neglected.
The only source of diamagnetism in this description are the nodal quasiparticles, whose contribution has been shown to be exceedingly small \cite{ghosal}. However, the mean field description does not include the \emph{direction fluctuations} of the bond currents themselves. Because fluctuating bond currents respond much more strongly than quasiparticles to an applied magnetic field (see below), it is possible that these fluctuations gives rise to an enhanced diamagnetic response.
 A suitable way to include these direction fluctuations is to formulate the DDW state in terms of a vertex model, in which the directed arrows represent directed bond
 currents (Fig.~1).
%(currents) on bonds on a square lattice can arrange themselves as a pattern of a set of vertices
%(one such arrangement being the ordered DDW state).
 In this paper we use this vertex model description of the DDW state to show that the diamagnetism of the state significantly enhances with increasing temperatures. Further, including also the magnitude fluctuations of the bond currents (not contained in the usual $6$-vertex model), which are important at high temperatures, we show that the high-$T$, low-field, $\chi$ can diverge as a power law of the applied field $H$. 
 \begin{comment}
 Our results are important in the context of the high-$T_c$ cuprates, but also, to our knowledge,
 this is the first study of the magnetic response of a particular vertex model, $6$-vertex model, which is an important statistical problem in itself.
 \end{comment}
%If these
% fluctuations are accounted for as in a vertex model, it is easy to see that the magnetic response of a bond current system is diamagnetic
%and can be large.

\paragraph{$6$-vertex model and F-model}
The classical vertex models were originally proposed to
study anti-ferroelectric materials and associated
phase transitions in electric fields \cite{rys,nagle}.
 One specific vertex model,
called the $6$-vertex model, is particularly interesting since it can be
 solved exactly by transfer matrices \cite{lieb1,lieb2,baxter}.
 The $6$-vertex model is defined by a set of vertices constructed out of directed arrow variables
 defined on the bonds of a square lattice. The arrows can represent any directed
  classical variable which serves as the building block of a thermodynamic statistical mechanical system.
 % For example, to describe the anti-ferroelectrics and the electric-field driven phase transition to
 %  a ferroelectric phase, the arrows are taken to represent local electric dipole moments.
  To describe an orbital current system
   the arrows are taken as directed bond currents.
 On the 2D $x-y$ plane, each of the nearest-neighbor
 bonds in the 4 directions $\bm d=\pm a\hat{x},\pm a\hat{y}$
from a vertex $v$ is associated with an orbital
 current  $I_{v}^{(\bm d)}$ of magnitude $I_0$. The current
  $I_{v}^{(\bm d)}$ is positive for current flowing
 parallel to $\bm d$ and negative for current flowing anti-parallel to
 $\bm d$.
In the steady state, there is no charge accumulation at each vertex and
 therefore the current is divergence-free $(\sum_{\bm d} I_{v}^{(\bm d)}=0)$.
This mandates that the total number of possible vertices on a
square lattice is $\frac{4!}{2!2!}=6$ (Fig.~1).
The  Hamiltonian describing the orientation of the currents for
a special case of the $6$-vertex model
called F model (\cite{rys,lieb1})is given by
\begin{equation}
H_0=\sum_{v,\bm d}-\frac{K}{2}(I_{v}^{(\bm d)}-I_{v}^{(-\bm d)})^2.
\label{eq:Fmodel0}\end{equation}
As is clear from Eq.~\ref{eq:Fmodel0}, in the F-model the anti-ferroelectric (AF) vertices
 (defined by  $I_v^{\hat x}=-I_v^{-\hat x}$ and $I_v^{\hat y}=-I_v^{-\hat y}$ ) are assigned
negative energies $-K$ and the rest of the vertices have energy $0$. Therefore, at low $T$,
 the ground state of the F-model is the ordered AF state,
which is nothing but the ordered DDW state when the arrows represent currents.  The
 AF state survives thermal fluctuations up to a critical temperature
 $T=T^*$. Above $T^*$ the current variables disorder
 into a critical phase
 with power-law current correlations \cite{baxter}.

\begin{figure}
\centering
\includegraphics[scale=0.35,angle=0]{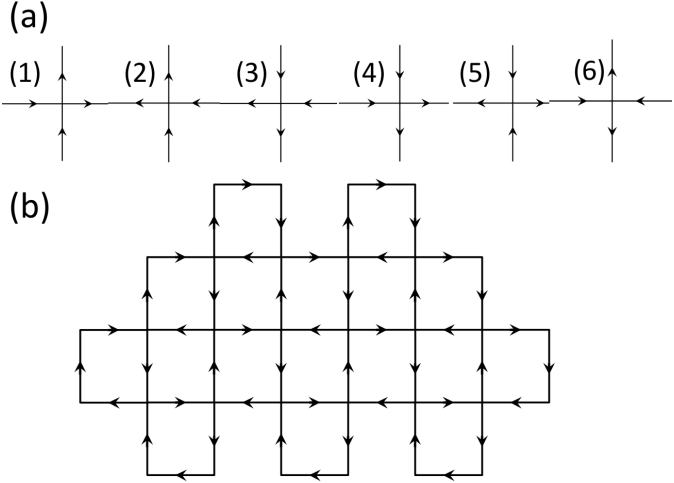}
\caption{(a) The six possible current vertices in the $6$-vertex model. The vertices (5) and (6) are the AF
vertices energetically favored by the F model. (b) The AF aligned low-temperature ground state
of the F model corresponds to the $d-$density wave (DDW) state proposed for the cuprate superconductors.
}
\end{figure}

\paragraph{DDW state and its relevance to the high-$T_c$ cuprates.}
 The singlet DDW state,
 described by an order parameter $ \left\langle \hat{c}_{\bm k+\bm Q,\alpha }^{\dagger }\hat{c}_{\bm k,\beta
}\right\rangle \propto iW_{\bm k}\,\delta _{\alpha \beta },\;W_{\bm k}=\frac{%
W_{0}}{2}(\cos k_{x}-\cos k_{y}) $, $\hat{c}_{\bm k}, \hat{c}_{\bm k}^{\dagger}$ are fermion operators, $\bm{k}$ is a 2D momentum,
 $\bm{Q}=(\pi, \pi)$, and $\alpha,\beta$ are spin indices, has been proposed as providing a phenomenologically consistent explanation for
the pseudogap phase of the underdoped cuprates \cite{Chakravarty01}. 
%According to this scenario, below the pseudogap temperature
%and below optimal doping the system develops  a non-zero value of $W_0$, which then evinces a $d$-wave gap (pseudogap).
  The assumption of DDW order below optimal
doping can lead to an explanation of numerous
experiments including the abrupt suppression of the superfluid density
 \cite{Tewari_superfluid}
and Hall number \cite{Tewari_hall} below
optimal doping as well as the more recent quantum oscillation experiments
 \cite{Sudip_oscillations} and Nernst effect \cite{Sumanta_nernst}.
 Mathematically, any Hamiltonian that leads to
$d$-wave superconductivity in the underdoped cuprates will
  almost certainly favor DDW order
as well \cite{Nayak,Nayak_Pivo}, making their coexistence and competition in the
phase diagram a plausible scenario.

      \paragraph{Connection of DDW state with F-model.}
      In a mean field picture, the only way the DDW state can thermally disorder is via a collapse of the magnitude of the order parameter  $W_0$
(i.e. collapse of the magnitudes of the currents themselves) at a second
 order thermal phase transition. However, this mean field description does
 not take into account the possible direction fluctuations of the bond
 currents.
 As is clear from Fig.~1,
 the ordered DDW state is nothing but the low-$T$ AF state of the F model.
In F model, with increase in $T$, the direction fluctuations
of the currents eventually make the system pass into a current disordered state
 above the  temperature $T^*$.
Thus the DDW state can thermally disorder by bond current
 fluctuations above $T^*$ long before the order parameter magnitude $W_0$ itself collapses
at a mean-field temperature $T_m>T^*$.
In this way the F-model and its quantum extension have recently been used \cite{sudip1,sudip2}
 to describe the thermal and quantum disordering transitions of the DDW state in the underdoped regime of the cuprates.

\paragraph{Diamagnetic response of F-model.} Let us first give an intuitive argument for the diamagnetism of the F-model.
The interaction of the orbital currents with
 an external magnetic field can be described by a term $H_{mag}=-\int d\bm r \bm J(\bm r)\cdot \bm A(\bm r)$ where
 $\bm J(\bm r)$ is the orbital current density and $\bm A(\bm r)$ is the vector potential. The expression for
$H_{mag}$ is derived by applying the minimal substitution $\bm p\rightarrow (\bm p- q \bm A/c)$ to the Schrodinger equation
for the electrons. The divergenceless orbital current $\bm J(\bm r)$ can be expressed in terms of a magnetization density $\bm m(\bm r)$,
 $\bm J(\bm r)=\bm\nabla\times \bm m(\bm r)$. Using the vector magnetic field
 $\bm B=\bm \nabla \times \bm A$ and integration by parts lead to the form
$H_{mag}=-\int d\bm r \bm J(\bm r)\cdot \bm A(\bm r)=\int d\bm r \bm m(\bm r)\cdot\bm B$.
 Thus, the bond currents can lower energy by aligning in circles perpendicular to $\bm B$ so that
 $\bm B\cdot \bm m(\bm r)<0$ (hence the response is diamagnetic).
 If such an arrangement of bond
 currents can be
accessed by flipping the local vertices in closed loops (to maintain the charge conservation)(Fig.~2), the resultant diamagnetic response
 can be large . Below we discuss this more quantitatively.

Adding the magnetic interaction term $H_{mag}$ to the Hamiltonian in Eq.~\ref{eq:Fmodel0} we get the total Hamiltonian as:
\begin{equation}
H=\sum_{v,\bm d}[-\frac{K}{2}(I_{v}^{(\bm d)}-I_{v}^{(-\bm d)})^2-a I_{v}^{(\bm d)}\bm d\cdot \bm A(v+\frac{\bm d}{2})].
\label{eq:Fmodel}
\end{equation}
Here $\bm A(x,y)$ is the vector potential given by $\bm A(x,y)=\frac{1}{2}(-B y, B x)$. The second term is equivalent to $H_{mag}$
when the current densities are limited to the bonds
\begin{figure}
\centering
\includegraphics[scale=0.4,angle=0]{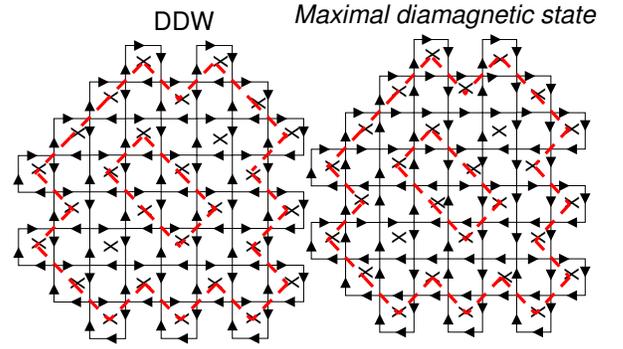}
\caption{(Color online) Two states, AF and the maximally current
 carrying state $l$ of the $6$-vertex model. The AF state
 is characterized by small clock-wise loops around plaquettes on one of the sublattices (marked by X).
To construct the maximally current carrying state $l$,
one starts with the AF state and
 reverses the counter-clockwise currents adjacent to all closed loops (red dashed curves). See text for details.
}\label{Fig2}
\end{figure}
and is equivalent to $\bm B \cdot\bm M$ where $\bm M$ is the total
magnetic moment.
To calculate the magnetization density we write the
partition function $Z=\sum_{j}e^{-E_j/k_B T}=e^{-E_0/k_B T}+\sum_{j\neq 0}e^{-E_j/k_B T}=e^{-E_0/k_B T}[1+\sum_{j\neq 0}e^{-(E_j-E_0)/k_B T}]$,
where $E_j$ is the
 energy associated with the configuration $j$ and $j=0$ is the minimum energy configuration.
 Using Eq.~\ref{eq:Fmodel}, the energy $E_j$ is given by
$E_j=E^v_j+\bm B \cdot \bm M_j$ where $E^v_j$ is the energy of the first term of $H$ and $\bm M_j$ is the total
magnetic moment of configuration $j$. Calling the configuration with the maximum diamagnetic moment $l$ (Fig.~\ref{Fig2} right panel),
 the free-energy $F=-k_B T\log{Z}$ can be written as
\begin{equation}
F=\bm B\cdot \bm M_l+\tilde{E}_0-k_B T \log{\left[1+\sum_{j\neq 0}e^{-\frac{(E_j-E_0)}{k_B T}}\right]}\label{eq:F}
\end{equation}
where $\tilde{E}_j=E^v_j+\bm B \cdot[\bm M_j-\bm M_l]$ and $\tilde{E}_0=\textrm{min}_j \tilde{E}_j$.
  Since $\tilde{E}_0\leq \tilde{E}_l=E^v_l$ and $\bm B\cdot\bm M_0\geq \bm B\cdot \bm M_l$,
 it follows that
 $E^v_0\leq \tilde{E}_0\leq E^v_l$. Additionally, the magnitude of the energy $E^v_j$ of any state $j$
 from Eq.~\ref{eq:Fmodel0} must be less than $K I_0^2 R^2$  where $R$ is the radius of the system containing $\sim R^2$ vertices.
 Thus the second term $\tilde{E}_0$ is bounded by $|\tilde{E}_0|\leq K I_0^2 R^2$.
The third term in Eq. ~\ref{eq:F} which is logarithmic also scales as $R^2$ since each of the terms under the summation over
 $j$ is less than unity and there are at most $6^{R^2}$ such terms corresponding to the current configurations on
$R^2$ vertices.
Combining these results, we find that
\begin{equation}
F=B M_l +\mathcal{O}(R^2)\label{eq4}
\end{equation} where $\mathcal{O}(R^2)$ represents corrections of order $R^2$ (which can be neglected as $|\bm M_l|\sim I_0 R^3/a$
as we show below).

 The state $l$ with maximum diamagnetic moment is understood by starting with the AF state as follows
 (Fig.~\ref{Fig2} ): Imagine large closed loops (red dashed curves) passing through
 the dual lattice points marked by the crosses.
  The currents on the bonds touching these loops but on the two opposite sides flow in opposite directions.
 For example, on the
  left panel of Fig. ~\ref{Fig2}, the bond currents right outside the loops are clockwise and right inside are counter-clockwise.
Reversing the counter-clockwise currents touching all such closed loops leads to the state $l$ shown on the
 right panel. The total magnetization $M$ is the product of the current and the total area enclosed
by all the clockwise loops. Since most of the clockwise loops  ($\sim R/a$ in number) enclose an area of order $R^2$, the total
magnetic moment is $\bm M_l\propto -\hat{z}I_0 R^3/a$ which is the desired result.
Using this equation for $M_l$ and neglecting terms of $\mathcal{O}(R^2)$ for large $R$
we get from Eqn.~\ref{eq4}, $F(B)=-B I_0 R^3/a$.
The magnetization density is calculated as $\bm m(B)=\frac{\hat{z}}{R^2}\frac{\partial F}{\partial B}$. This gives
$m(B)\sim -\hat{z} I_0 R/a$, which is divergent in the thermodynamic limit $(R\rightarrow \infty)$ for any
 non-vanishing $B$.

So far we have ignored the magnetic field generated by the induced currents themselves. Such
 a field results in a magnetostatic current-current interaction.
 The current-current interaction can be accounted for by using simple
  magnetostatics,
\begin{equation}H=B+4\pi |\bm m(B)|,\label{eq:mag}
\end{equation} where $B$ is the magnetic field
 and $H$ is the magnetic induction which can be taken as
the externally applied magnetic field.
 Here $\bm m(B)$ is the magnetization density which is
 opposite in direction to B (diamagnetic)  in sign and
 increases in magnitude from $|\bm m(B)|=0$ at
$B=0$ to $|\bm m(B)|\sim I_0 R/a$ for any magnetic field
$B\gtrsim \mathcal{O}(1/R)$. To estimate the solution $B$ of Eq.~\ref{eq:mag}
 let us define the function $f(B)=B+4\pi |\bm m(B)|-H$ such that Eq.~\ref{eq:mag}
 is written as $f(B)=0$. Since $H < I_0 R/a$ ($R\rightarrow \infty$ in the thermodynamic limit), it follows that
 $f(B)>0$ for $B\gtrsim \mathcal{O}(1/R)$. On the other hand $f(B=0)=-H<0$. Therefore $f(B)$, being an increasing
function of $B$, has a unique root satisfying the constraint
 $B\lesssim \mathcal{O}(1/R)$.  Since in the thermodynamic limit
$R\rightarrow \infty$, it follows that $B$ is completely
 expelled from the system and it behaves like a perfect diamagnet
 similar to a type I superconductor.

\paragraph{Diamagnetism in the AF phase.}
Despite the above analysis, the low-$T$ ($T\ll T^*$) AF phase being gapped is not
 expected to have a large diamagnetic response. The response of the
 AF phase to a magnetic field should be dominated by
 flips of small current loops. The combination of these elementary
current loop flips can generate a flip of a large loop of
length $L$ which has a magnetic moment $|\bm M|\sim I_0 L^2$.
From Eq.~\ref{eq:Fmodel0} such loop flips cost  energy $K L$ (flipping each AF vertex costs energy $K$ and $L$ such vertices need to be flipped
for a loop of length $L$). However, the applied field lowers the
energy of such a current loop by $-B |\bm M|=-B I_0 L^2$.
Thus the energy cost of a flipped loop, $V(L)=K L- B I_0 L^2$, is
 positive for small $L$, has a positive peak at $L=K/2 B I_0$ and
 becomes negative for large $L$.
Thus only loops that form out of a thermal fluctuation with an energy larger than $\textrm{max}_L V(L)=K^2/2 B I_0$
 can cross the threshold value of $L=K/2 B I_0$ to become a large loop.
The fraction of such high energy loops is determined by the Boltzmann factor as $e^{-K^2/2 k_B T B I_0}$.
  Thus the AF state is stable for low-$T$ and $B$ (i.e $e^{-K^2/2 k_B T B I_0}\ll 1$). This conclusion is also consistent with the numerical monte-carlo
 simulations \cite{kineticbcsos}. It also follows
that as $T$ increases ($T$ is a substantial fraction of $T^*$),
the diamagnetic response of the AF state should increase significantly.
However, the time-scale for development of diamagnetic response is
 expected to become longer as
temperature becomes smaller than $T^*$ leading to possible hysteretic
 behavior of the magnetization
as a function of applied magnetic field.  
%The width of histeretic loops should depend on effects
%which determine the kinematics of the current loops. These are  the
 % height of the energy barriers (given by $K^2/2 B I_0$),  cooling rate, impurity pinning etc.

\paragraph{Diamagnetism in the critical phase.}
For $T>T^*$, the AF order is completely destroyed and the system develops critical
 current fluctuations whose correlation
is scale-invariant. From the argument in the previous paragraph it
follows that this phase is strongly diamagnetic.
 The scale invariance of the fluctuations in the critical phase
allows one to describe the
critical phase by a continuum theory such as a height model
 \cite{beijeren,henley}.
In the height model the magnetization density $\bm m(\bm r)$ is mapped
 to the vertical displacement $h(\bm r)$ of a $2D$ surface
such that $\bm m(\bm r)= I_0 h(\bm r)\hat z$ and
 $\bm J(\bm r)=\bm \nabla\times \bm m(\bm r)$.

The current-current correlations in the high $T$ critical phase of the
 F model
are obtained from the Gaussian theory of height
 fluctuations \cite{beijeren,henley} described by
the coarse-grained continuum Hamiltonian
 \begin{equation}
H=\int d^2\bm r \tilde{K}|\nabla h|^2- \bm B\cdot \bm M_{tot}\label{eq:criticalh}
\end{equation}
where $\bm M_{tot}$ as before is the total magnetization of the model.

From the argument in the last section, it is clear that for
 $T\gtrsim T^*$, the F model responds to a magnetic field by generating
large  current loops. This results in the formation of
patches of circulating currents.
% (of radius, say, $R_p$).
If the sizes of these patches are macroscopically large,
 this would lead to perfect diamagnetism. However, so far we have
neglected the
% \textbf{spontaneous (thermal)}
\textit{magnitude} fluctuations of the bond currents $I_0$, which
should be taken into account at high $T$.
% (magnitude fluctuation of the
%currents results from fluctuations in the magnitude of $W_0$).
Such magnitude fluctuations of $I_0$
can occur from spontaneous thermal fluctuations of the DDW gap magnitude $W_0$ and variations of the
local density of quasiparticles.
%of the local density of quasiparticles in this phase.
Here by quasiparticles we mean the quasiparticles in the DDW state that carry charge \cite{sudip3}.
In the presence of such magnitude fluctuations the bond-current
$I_0$ is allowed to vary spatially as $I_0(\bm r,t)$ so that the
current density  is now given by $\bm J(\bm r,t)=I_0(\bm r,t)\bm \nabla \times (h(\bm r)\hat{\bm z})$.
Therefore it is no longer a divergence-free quantity.
%Here we neglect possible couplings between the fermionic
%quasiparticles with the current fluctuations themselves.
The magnitude fluctuations of the currents
 will lead to a cut-off length-scale $R_0>R_p$ for the patch sizes $R_p$.
The introduction of this  cut-off length-scale in the original $F$
model directly leads to a power-law dependence of $m$ on $B$ as we show at the end of this section.
Below we first speculate on a possible mechanism for the emergence of this cut-off scale.

Let us consider a patch with a circularly symmetric current density
$\bm J(\bm r,t)=I_0(\bm r,t)(\bm \nabla h(r)\times \hat{\bm z})=I_0(\bm r,t)h'(r)\hat{\bm \theta}$ corresponding
to a circularly symmetric height profile $h(r)$.
% In systems with a
%finite quasiparticle density (i.e., high $T$),
Apart from a thermal fluctuation component, $\zeta(\bm r,t)$, the magnitude of the local current $I_0(\bm r,t)$  depends on the
 density of quasiparticles $n(\bm r,t)$.  Writing $I_0(\bm r,t)=I_0 (S(n(\bm r,t))+\zeta(\bm r,t))$, the time-varying
current density $\bm J(\bm r,t)$ is now given by $\bm J(\bm r,t)=I_0(\bm r,t)h'(r)\hat{\bm \theta}=I_0 h'(r)[S(n(\bm r,t)+\hat{\bm\theta}\cdot\bm\nabla\zeta(\bm r,t)]\hat{\bm \theta}=J(r)[S(n(\bm r,t))+\hat{\bm\theta}\cdot\bm\nabla\zeta(\bm r,t)]\hat{\bm \theta}$,
 where $h'(r)=\frac{dh(r)}{dr}$, $\zeta(\bm r,t)$ is the noise term that accounts for the
spontaneous thermal fluctuations of the current density and $\hat{\bm\theta}$ is the tangential direction around a loop on the circular patch.
Since charge density is conserved we have the continuity equation,
%$\partial_t n(\bm r,t)+J(r)\bm\nabla\cdot\left\{\hat{\bm\theta}S(n(\bm r,t))\right\}=D \nabla^2 n$,
$\partial_t n(\bm r,t)+\bm\nabla\cdot\bm J(n(\bm r,t))=D \nabla^2 n$,
 where $D$ is the diffusion constant of the quasiparticles.
 For a circular profile of a single patch of circulating currents, this equation can be
% decoupled
% into a set of 1D conservation law equations for each radius $|\bm r|=r$ and
 written as
\begin{equation}
\partial_t n+J(r)[S'(n)\hat{\bm \theta}\cdot\bm\nabla n+\hat{\bm \theta}\cdot\bm\nabla\zeta(\bm r,t)]=D \nabla^2 n.\label{conservation}
\end{equation}

The second term in the above equation, which is referred to as the bond-current term drives current only along
the tangential direction to the loop at a given radius and hence is one-dimensional in character. The
 quasiparticle diffusion term on the right hand side of the above equation is two dimensional in space and
in general will have a finite radial component. However, long wave-length fluctuations in the bond current
$\zeta(\bm r,t)$ (the correlation function of $\zeta(\bm r,t)$ is taken as $\expect{\zeta(\bm r,t)\zeta(\bm r',t')}\sim \delta(\bm r-\bm r')\delta(t-t')$), with a length-scale $\lambda$, create long wave-length variations in the quasiparticle
 density which are tangential in direction (as shown by the third term in Eq.~\ref{conservation}).
 The evolution of the quasiparticle density following such a fluctuation,
which is described by Eq.~\ref{conservation}, is dominated by the tangential bond-current term (i.e. second term),
 which scales as
$\lambda^{-1}$, and the contribution of the radial diffusion term (which scales as $\lambda^{-2}$) is subdominant
for large $\lambda$. Therefore, the long length-scale behavior of Eq.~\ref{conservation} can be understood in
terms of approximately decoupled 1D conservation law equations for each loop at radii $|\bm r|=r$.
As shown in Refs.~\cite{whitham,beijeren1,jensen}, for mean-field bond-current densities $S(n)$ which
vanish for very small or large quasiparticle densities, such 1D equations are unstable to the proliferation
of long-lived shock solutions  and the current density $I_0$ is expected to drop to zero at some radius $R_0$ which can be determined
by solving the relation $J(r=R_0)\propto R_0^{-\alpha}$. In the present case, how such a relation arises
can be seen in the following way.
From Eq.~\ref{conservation}, we see that the bond-current term at a length-scale $R_0$ scales as $J(r=R_0)R_0^{-1}$,
while the diffusion term scales as $D R_0^{-2}$.
%for instabilities
%from the largest length-scale radius $R_0$,  proportional to $J(r=R_0)$,
Thus the bond-current term, which drives the instability,  dominates at a radius $R_0$
when $J(r=R_0)R_0^{-1}\sim D R_0^{-2}$. This leads to the constraint
 $J(r=R_0)\propto R_0^{-\alpha}$ with $\alpha=1$.
In reality this simple argument ignores the $R_0$-dependence of $n_s$ \cite{whitham,fogedby}, which leads to a value of
$\alpha$ slightly greater than 1.
%The simplistic analysis presented ignores the length-scale dependence
%of the magnitude of the fluctuations in $n(x,t)$ that lead to the solitonic instability \cite{whitham}. For our
%purposes, we will assume that this renormalizes $\alpha$ slightly so that $\alpha>1$.

Let us now return to the F model and show that the introduction
of a cut-off scale leads to a power-law dependence of $m$ on $B$.
We consider the free-energy of a patch of finite radius $R_0$.
% Since the current density $J(R_0)$ decays with $R_0$
%we assume $J(R_0)\propto I_0 h'(R_0)\propto R_0^{-\alpha}$ where $\alpha>0$ to ensure
%an inverse relation between the cut-off current and radius.
%Considering the evolution of random charge density fluctuations $n$ under the Burger's equation (i.e setting $S[n]=J_0 n^2$ in the
%conservation equation) leads to an exponent $\alpha=3/2$.
%Any other choice would lead to qualitatively similar behavior as below.
Using Eq.~\ref{eq:criticalh} the free-energy density of such a patch is
 given by
 \begin{equation}
f=\int_0^{R_0}\frac{2 r d r}{R_0^2} (\tilde{K} h'^2-I_0 B h)\label{eq:hnew}.
\end{equation}
Minimizing $f$ with respect to $h$ using the
usual method of variations (and without any approximations)
 leads to $h'(r)=-B I_0 r/4\tilde{K}$.
 Using the relation $I_0 h'(R_0)=R_0^{-\alpha}$, we get
 $R_0\propto B^{-1/(1+\alpha)} $. Substituting this in Eq.~\ref{eq:hnew} leads to $f\propto B^{2}R_0^2\sim B^{2\alpha/(1+\alpha)}$.
 The diamagnetic susceptibility $\chi=\frac{\partial^2 f}{\partial B^2}\propto B^{-2/(1+\alpha)}$
 diverges as an inverse power-law.

\paragraph{Conclusion.}
We analyze the diamagnetic response of the $6$-vertex model, which is
used to model the DDW phase proposed for the pseudo-gap phase of
the high-$T_c$ cuprates. We find that
deep in the low-$T$ AF phase the diamagnetic response is weak.
 With increasing $T$, especially for $T \lsim T^*$, the diamagnetism is
significantly enhanced. The disordered critical phase
 for $T > T^*$, is perfectly diamagnetic within the strict
 $6$-vertex model.  With the magnitude
fluctuations of the current (magnitude fluctuations of the DDW
order parameter $W_0$) taken into account, the low-field
 diamagnetic susceptibility
$\chi$ in this phase diverges as a power-law of the field.

%%%%%%%%%%%%%%%%%%%%%%%%%%%%%%%%%%%%%%%%%%%%%%%%%%%%%%%%%%%%%%%%%%%%%

We thank S. Chakravarty for pointing out a crucial error in an earlier version of the manuscript.
We also thank J. Toner and N. P. Armitage for valuable discussions.
 We acknowledge the hospitality of the Aspen Center for Physics where part of this work was completed. Sau acknowledges DARPA-QuEST and JQI-NSF-PFC, and Tewari acknowledges
DARPA MTO Grant No:FA9550-10-1-0497 for support.
%ST acknowledges DOE/EPSCoR Grant \# DE-FG02-04ER-46139 and Clemson University start up funds for support.

%%%%%%%%%%%%%%%%%%%%%%%%%%%%%%%%%%%%%%%%%%%%%%%%%%%%%%%%%%%%%%%%%%%%%
%%%%%%%%%%%%%%%%%%%%%%%%%%%%%%%%%%%%%%%%%%%%%%%%%%%%%%%%%%%%%%%%%%%%%

%\bibliography{./refs_pwave_top_ins}

%\end{document}

%%%%%%%%%%%%%%%%%%%%%%%%%%%%%%%%%%%%%%%%%%%%%%%%%%%%%%%%%%%%%%%

\end{document}